\documentclass[a4paper,oneside]{article}
\usepackage[english]{babel}
\usepackage{amsmath}
\usepackage{graphicx}
\usepackage[colorlinks=true, allcolors=blue]{hyperref}
\usepackage[utf8]{inputenc}
\usepackage{amssymb}
\usepackage{amsfonts}
\usepackage[warn]{mathtext}
\usepackage{float}
\usepackage{layout}
\usepackage{revsymb}
\usepackage{multirow}
\usepackage{authblk}
\title{Generation of narrow beams of ultrarelativistic positrons (electrons) in the resonant strong electromagnetic field-assisted Breit-Wheeler  process}
\author[1]{S.P. Roshchupkin}
\author[1]{V.D. Serov}
\author[1]{V.V. Dubov}
\date{}
\affil[1]{Peter the Great St. Petersburg Polytechnic University,
195251, St-Petersburg, Russian Federation, Russia}
\begin{document}
\maketitle
\begin{abstract}
    The resonant external field-assisted Breit-Wheeler process (Oleinik resonances) for strong electromagnetic fields with intensities less than the critical Schwinger field has been theoretically studied. The resonant kinematics has been studied in detail. The case of high-energy initial gamma quanta and emerging ultrarelativistic electron-positron pairs is studied. The resonant differential cross section is obtained. The generation of narrow beams of ultrarelativistic positrons (for Channel A) and electrons (for Channel B) is predicted with a probability significantly exceeding corresponding to the non-resonant process.
\end{abstract}
\section{Introduction}
\par Over the past several decades, there has been significant interest in studying the processes of quantum electrodynamics (QED) in external electromagnetic fields (see, for example, reviews \cite{1}-\cite{7}, monographs \cite{8}-\cite{10} and articles \cite{11}-\cite{55}). This is mainly associated with the appearance of lasers with high radiation intensities and beams of small transverse dimensions \cite{11}-\cite{19}.
\par An important feature of high-order by the fine structure constant QED processes in an external field is the potential for their resonant occurrence, where virtual intermediate particles enter the mass shell. Such resonances were first considered by Oleinik \cite{20,21}. Under resonance conditions, the conservation laws of energy and momentum are satisfied for intermediate particles in an external field. As a result, second-order processes by the fine structure constant effectively reduce into two sequential first-order processes. A detailed discussion of resonant processes is presented in reviews \cite{2,4}, monographs \cite{8,9,10}, as well as recent articles \cite{31}-\cite{36}. It is important to note that the probability of resonant processes can significantly exceed the corresponding probabilities of non-resonant processes.
\par The process of electron-positron pair production by two gamma quanta was first considered by Breit and Wheeler \cite{37}. Currently, there is a significant number of works devoted to the study of the Breit-Wheeler process in an external electromagnetic field (see, for example, \cite{38}-\cite{49}). It should be noted that a distinction should be made between the external field-stimulated Breit-Wheeler process (a first-order process with respect to the fine structure constant) and the external field-assisted Breit-Wheeler process (a second-order process with respect to the fine structure constant). In this paper, Oleinik's resonances for the external strong field-assisted Breit-Wheeler process will be investigated. It should be noted that in a weak field, this process was considered in the article \cite{49}. It is important to note that under the conditions of resonance and the absence of interference between different reaction channels, the original second-order process effectively reduces to two first-order processes: the external field-stimulated Breit-Wheeler process and the external field-stimulated Compton effect \cite{49}.
\par The main parameter for describing the Breit-Wheeler process in the field of a plane electromagnetic wave is the classical relativistic-invariant parameter
\begin{equation}\label{1}
\eta=\frac{eF\lambdabar}{mc^2},
\end{equation}
 numerically equal to the ratio of the work of the field on the wavelength to the rest energy of the electron. Here $e$ and $m$ are the charge and mass of the electron, $F$ and $\lambdabar=c/\omega$ are the electric field strength and wavelength, $\omega$ is the frequency of the wave \cite{1}.
\par In this paper, we consider the resonant strong electromagnetic field-assisted Breit-Wheeler process for high-energy gamma quanta with energies $\hbar\omega_{1,2}\lesssim10^2$ GeV. Therefore, we will consider high-energy gamma quanta in the following, ensuring that the produced electron-positron pair in a field of the wave is ultrarelativistic
\begin{equation}\label{2}
\hbar\omega_{1,2}\gg mc^2,\quad E_{\pm}\gg mc^2.
\end{equation}
Here $\hbar\omega_{1,2}$ and $E_{\pm}$ are energies of the initial gamma quanta and final positron or electron. Therefore, we will assume that the magnitude of the classical parameter $\eta$ is upper bounded by the condition:
\begin{equation}\label{3}
\eta\ll\eta_{max},\quad \eta_{max}=\mathrm{min}\left(\frac{E_{\pm}}{mc^2}\right).
\end{equation}
\par Let's estimate the maximum intensity of the electric field in the wave. For electron-positron pair energies $E_{\pm}\lesssim10^2$ GeV,  it follows from equation (\ref{3}) that $\eta\ll\eta_{max}\sim10^5$, or for the field strength we have $F\ll F_{max}\sim10^{15}$ Vcm$^{-1}$ ($I\ll I_{\max}\sim10^{28}$ Wcm$^{-2}$). Thus, the problem will consider sufficiently large intensities of the electromagnetic wave. However, these fields must be smaller than the Schwinger critical field $F_*\approx1.3\cdot10^{16}$ Vcm$^{-1}$ \cite{5,52}.
\par In the following, the relativistic system of units is used: $c=\hbar=1$.

\section{Amplitude of the process}
\par
\begin{figure}[H]
    \centering
\begin{minipage}{.49\textwidth}
    \centering
\includegraphics[width=0.8\linewidth]{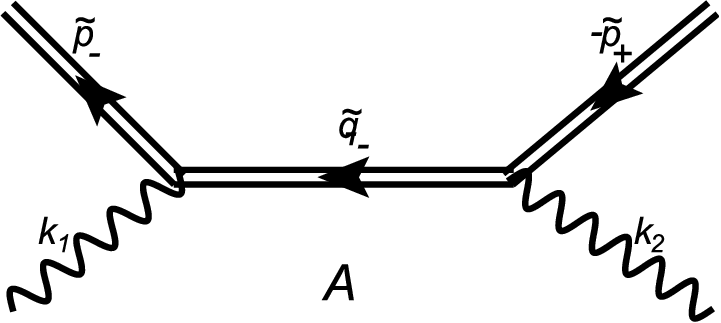}
\end{minipage}
\hfill
\begin{minipage}{.49\textwidth}
    \centering
   \includegraphics[width=0.8\linewidth]{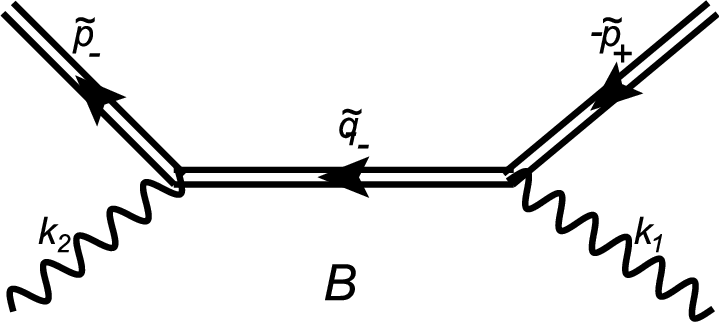}
\end{minipage}
\caption{Feynman diagrams of  electron--positron-pair production by two gamma quanta in an external field. External double lines correspond to Volkov functions of electron or positron, wavy lines correspond to wave functions of initial gamma quanta, and internal double lines correspond to Green's electron function in the field of a plane electromagnetic wave.}
\label{fig1}
\end{figure}

 Let us consider this process in the field of a plane circularly
polarized wave propagating along the z axis:
\begin{equation}\label{4}
A(\varphi)=\frac{F}{\omega}\left(e_x\cos{\varphi}+\delta e_y\sin{\varphi}\right), \quad \varphi=(kx)=\omega(t-z),\quad\delta=\pm1.
\end{equation}
Here $e_x, e_y$ are the polarization 4-vectors of the external field that have the following properties:
$e_x=(0,\mathbf{e_x}), \quad e_y=(0,\mathbf{e_y}), \quad e_xe_y=0, \quad (e_x)^2=(e_y)^2=-1.$ The external field-assisted Breit-Wheeler process is characterized by two Feynman diagrams (Fig.\ref{fig1}). 
\par The amplitude of the considered process is written as follows
\begin{equation}\label{5}
S_{if}=ie^2\iint d^4x_1d^4x_2\overline{\Psi}_{p_-}(x_1|A)\hat{A}_1(x_1;k_1)G(x_2x_1|A)\hat{A}_2(x_2;k_2)\Psi_{-p_+}(x_2|A)\, +\, (k_1\leftrightarrow k_2),
\end{equation}
where $k_{1,2}=(\omega_{1,2},\mathbf{k_{1,2}})$ --- 4-momenta of the initial gamma quanta, $p_{\pm}=(E_{\pm}, \mathbf{p_{\pm}})$ --- 4-momenta of the final electron and positron. Here and further, the notation for the convolution of a 4-vector with the Dirac gamma matrices is used: $\hat{A}_{1,2}\equiv\gamma_{\mu}A^{\mu}_{1,2}\quad\mu=0,1,2,3$. The 4-potentials of the initial gamma quanta $A_j$ in expression (\ref{5}) are determined by the functions
\begin{equation}\label{6}
A_j(x;k_j)=\sqrt{\frac{2\pi}{\omega_j}}\varepsilon_je^{-ik_jx}, \quad j=1,2,
\end{equation}
where $\varepsilon_j$ --- 4-vectors of the polarization of the initial gamma quanta.
\par In the amplitude (\ref{5}), the electron-positron pair corresponds to the Volkov functions \cite{50,51}:
\begin{equation}\label{7}
\Psi_{p}(x|A)=\mathfrak{J}_{p}(x)\frac{u_p}{\sqrt{2E}}, \quad\mathfrak{J}_{p}(x)=\left[1+\frac{e}{2(pk)}\hat{k}\hat{A}(kx)\right]e^{iS_{p}(x)},
\end{equation}
\begin{equation}\label{8}
S_{p}(x)=-(px)-\frac{e}{(kp)}\int_0^{kx}d\varphi[pA(\varphi)-\frac{e}{2}A^2(\varphi)],
\end{equation}
where $u_p$ is the Dirac bispinor. The intermediate state in the amplitude (\ref{5}) corresponds to the Green's function of the electron in the field of a plane wave $G(x_2x_1|A)$ \cite{53}:
\begin{equation}\label{9}
G(x_2x_1|A)=\int\frac{d^4p}{(2\pi)^4}\mathfrak{J}_{p}(x_2)\frac{\hat{p}+ m}{p^2-m^2}\overline{\mathfrak{J}}_{p}(x_1).
\end{equation}
\par After simple transformations, the amplitude (\ref{5}) can be represented as follows:
\begin{equation}\label{10}
S_{if}=\sum_{l=-\infty}^{+\infty}S_l,
\end{equation}
where the partial amplitude $S_l$ corresponds to the absorption or emission of  $|l|$ photons of the external wave. For the Channel A, the partial amplitude can be represented in the following form:
\begin{equation}\label{11} 
S_l=\frac{i\pi e^2(2\pi)^4e^{-id}}{\sqrt{\widetilde{E}_-\widetilde{E}_+\omega_1\omega_2}}\left[u_{p_-}M_{l}v_{p_+}\right]\delta^{(4)}\left(k_1+k_2-\widetilde{p}_--\widetilde{p}_+-lk\right).
\end{equation}
Here $d$  is the phase, independent of the summation indices, $M_l$ --- the matrix determined by the expression
\begin{equation}\label{12}
M_l=\varepsilon_{1\mu}\varepsilon_{2\nu}\sum_{r=-\infty}^{+\infty}K^{\mu}_{l+r}(\widetilde{p}_-,\widetilde{q}_-)\frac{\hat{q}_-+m}{\widetilde{q}_-^2-m_*^2}K^{\nu}_{-r}(\widetilde{q}_-,-\widetilde{p}_+),\quad\mu,\nu=0,1,2,3.
\end{equation}
In relation (\ref{12}), the functions $K^{\mu}_{l+r}$ and $K^{\nu}_{-r}$ have the following form:
\begin{equation}\label{13}
K^{\mu'}_{n}(\widetilde{p}',\widetilde{p})=a^{\mu'}L_n(\widetilde{p}',\widetilde{p})+b_-^{\mu'}L_{n-1}+b_+^{\mu'}L_{n+1}.
\end{equation}
Here, the matrices $a^{\mu'}$ and $b^{\mu'}_{\pm}$ have the following form:
\begin{equation}\label{14}
a^{\mu'}=\gamma^{\mu'}+\frac{m^2\hat{k}}{2(k\widetilde{p'})(k\widetilde{p})}k^{\nu},\quad b^{\mu'}_{\pm}=\frac{1}{4}\eta m\left(\frac{\hat{e}_{\pm}\hat{k}\gamma^{\mu'}}{(k\widetilde{p'})}+\frac{\gamma^{\mu'}\hat{k}\hat{e}_{\pm}}{(k\widetilde{p})}\right),
\end{equation}
\begin{equation}\label{15}
    e_{\pm}\equiv e_x\pm ie_y,\quad\mu'=\mu,\nu,\quad n=l+r,-r,\quad\widetilde{p}=-\widetilde{p}_+,\widetilde{q}_-,\quad\widetilde{p'}=\widetilde{q}_-,\widetilde{p}_-.
\end{equation}
In relations (\ref{12}), (\ref{13}) there are special functions $L_n$ \cite{3}, which in the case of circular polarization of the wave can be represented using Bessel functions with integer indices
\begin{equation}\label{16}
L_n(\widetilde{p}',\widetilde{p})=\exp(-in\chi_{\widetilde{p}'\widetilde{p}})J_n(\gamma_{\widetilde{p}'\widetilde{p}}),
\end{equation}
where is denoted
\begin{equation}\label{17}
\gamma_{\widetilde{p}'\widetilde{p}}=m\eta\sqrt{-Q^2_{\widetilde{p}'\widetilde{p}}}, \quad\tan\chi_{\widetilde{p}'\widetilde{p}}=\delta\frac{(Q_{\widetilde{p}'\widetilde{p}}e_y)}{(Q_{\widetilde{p}'\widetilde{p}}e_x)}, \quad Q_{\widetilde{p}'\widetilde{p}}=\frac{\widetilde{p}'}{(p'k)}-\frac{\widetilde{p}}{(pk)}.
\end{equation}
\par In the expressions (\ref{11}) and (\ref{12}) $\widetilde{p}_{\pm}=(\widetilde{E}_{\pm}, \mathbf{\widetilde{p}_{\pm}})$ and $\widetilde{q}_-$ are the 4-quasimomenta of the electron (positron) and intermediate electron, $m_*$ is the effective mass of the electron in the
field of a circularly polarized wave (\ref{4}) \cite{34}:
\begin{equation}\label{18}
\widetilde{q}_-=k_2+rk-\widetilde{p}_{\pm},
\end{equation}
\begin{equation}\label{19}
\widetilde{p}_{\pm}=p_{\pm}+\eta^2\frac{m^2}{2(kp_{\pm})}k,\quad\widetilde{q}_-=q_-+\eta^2\frac{m^2}{2(kq_-)}k,
\end{equation}
\begin{equation}
    \widetilde{p}_{\pm}^2=m_*^2,\quad m_*=m\sqrt{1+\eta^2}.
\end{equation}
\section{The resonant kinematics}
\par
\begin{figure}[H]
    \centering
\begin{minipage}{.49\textwidth}
    \centering
   \includegraphics[width=0.8\linewidth]{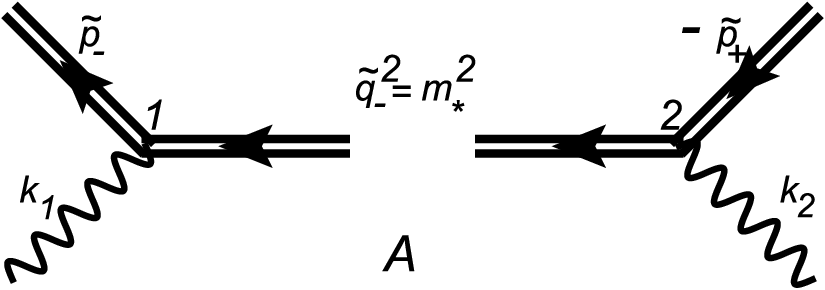}
\end{minipage}
\hfill
\begin{minipage}{.49\textwidth}
    \centering
   \includegraphics[width=0.8\linewidth]{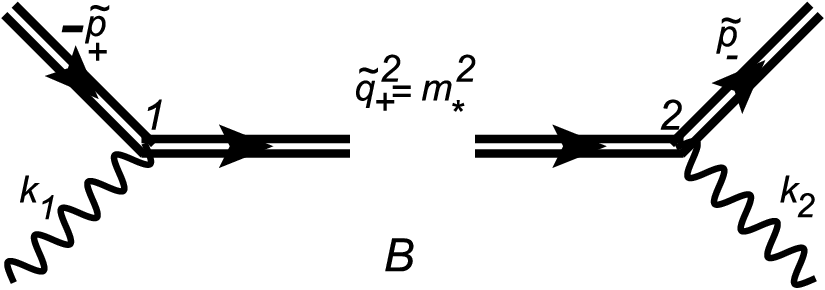}
\end{minipage}
\caption{Feynman diagrams of the resonant electron–positron-pair production by two gamma quanta in an external
field, Channels A and B; for Channels A' and B' $k_1\leftrightarrow k_2$.}
\label{fig2}
\end{figure}

 Under resonance conditions, both an electron and a positron can be intermediate particles. Therefore, instead of two Feynman diagrams in the non-resonant case (see Fig. \ref{fig1}), under resonance conditions we will have 4 Feynman diagrams (see Fig. \ref{fig2}):  Channels A and B, as well as Channels A' and B', which are obtained from channels A and B by rearranging the initial gamma quanta $(k_1\leftrightarrow k_2)$. Each channel in the resonance conditions effectively decays into two first-order processes by the fine structure constant: the external field-stimulated Breit-Wheeler process (EFSBWP) and the external field-stimulated Compton effect (EFSCE) with intermediate electrons and positrons entering the mass shell:
\begin{equation}\label{21}
\widetilde{q}_-^2=m_*^2,\quad\widetilde{q}_+^2=m_*^2.
\end{equation}
\par Further consideration will be carried out for resonant Channels A and B (see Fig. \ref{fig2}). It is important to emphasize that the laws of conservation of energy-momentum for intermediate processes of resonant Channels A and B have the form:
\begin{equation}\label{22}
\mathrm{EFSBWP}:\qquad k_2+rk=\widetilde{q}_{\mp}+\widetilde{p}_{\pm}\quad r=1,2,3\ldots;
\end{equation}
\begin{equation}\label{23}
    \mathrm{EFSCE}:\qquad k_1+\widetilde{q}_{\mp}=\widetilde{p}_{\mp}+r'k\quad r'=1,2,3\ldots\quad(r'=l+r).
\end{equation}
\par Since the problem considers high-energy initial gamma quanta and ultrarelativistic energies of the final electron-positron pair (\ref{2}), under such conditions, the momenta of the initial and final particles should lie within a narrow angle cone, which should be far away from the direction of wave propagation:
\begin{equation}\label{24}
\theta_{j\pm}\equiv\angle(\mathbf{k}_j, \mathbf{p}_{\pm})\ll1,\quad\theta_{i}\equiv\angle(\mathbf{k}_1, \mathbf{k}_2)\ll1,
\end{equation}
\begin{equation}\label{24'}
\theta\equiv\angle(\mathbf{p}_{\pm},\mathbf{k})\sim1,\quad\theta_j\equiv\angle(\mathbf{k}_j, \mathbf{k})\sim1,\quad j=1,2;\quad\theta\approx\theta_1\approx\theta_2.
\end{equation}
\par Let us note that under conditions (\ref{2}), (\ref{3}), the expression for the positron (electron) quasienergy can be simplified:
\begin{equation}\label{25}
\widetilde{E}_{\pm}=E_{\pm}\left[1+\frac{1}{4\sin^2\frac{\theta_{\pm}}{2}}\left(\frac{m\eta}{E_{\pm}}\right)^2\right]\approx E_{\pm}.
\end{equation}
\par Let us determine the resonance energy of the positron (electron) for the second vertex (see Fig. \ref{fig2}). Taking into account relations (\ref{2}), (\ref{3}), (\ref{21}), (\ref{24}) from the conservation of 4-momentum law (\ref{22}) for the external field-stimulated Breit-Wheeler process, we obtain the resonance energies of the positron  (for Channel A) or electron  (for Channel B) in units of the total energy of the initial gamma quanta:
\begin{equation}\label{26}
    x_{j'(r)}=\frac{\omega_2}{2\omega_i(\varepsilon_{2BW(r)}+\delta^2_{2j'})}\left[\varepsilon_{2BW(r)}\pm\sqrt{\varepsilon_{2BW(r)}(\varepsilon_{2BW(r)}-1)-\delta^2_{2j'}}\right],\quad j'=+,-.
\end{equation}
Here it is indicated:
\begin{equation}\label{27}
    x_{\pm(r)}=\frac{E_{\pm}(r)}{\omega_i},\quad\omega_i=\omega_1+\omega_2,\quad\delta_{2\pm}=\frac{\omega_2}{2m_*}\theta_{2\pm}.
\end{equation}
In this case, the ultrarelativistic parameter $\delta_{2\pm}$, which determines the outgoing angle of the positron or electron, is contained within the interval 
\begin{equation}\label{28}
0\le\delta^2_{2+}\le\delta^2_{2+max},\quad\delta^2_{2+max}=\varepsilon_{2BW(r)}(\varepsilon_{2BW(r)}-1).
\end{equation}
It is important to emphasize that in equation (\ref{26}), the quantity $\varepsilon_{2BW(r)}$ is bounded from below by unity
\begin{equation}\label{29}
\varepsilon_{2BW(r)}=r\varepsilon_{2BW}\ge1,\quad\varepsilon_{2BW}=\frac{\omega_2}{\omega_{BW}},
\end{equation}
where $\omega_{BW}$ is the characteristic quantum energy of the external field-stimulated Breit-Wheeler process:
\begin{equation}\label{30}
   \omega_{BW}=\frac{m_*^2}{\omega\sin^2\frac{\theta}{2}}=\left\{ \begin{array}{rcl}
174\mbox{GeV}&\mbox{if}&\omega=3\mbox{eV},I=1.675\cdot10^{19}\mbox{Wcm}^{-2}\\
5.22\mbox{GeV}&\mbox{if}&\omega=0.1\mbox{keV},I=1.861\cdot10^{22}\mbox{Wcm}^{-2}\\
52.2\mbox{MeV}&\mbox{if}&\omega=10\mbox{keV},I=1.861\cdot10^{26}\mbox{Wcm}^{-2}
   \end{array}\right.
\end{equation}
When estimating the value of the characteristic energy, frequencies of electromagnetic waves in the optical and X-ray ranges were used in equation (\ref{30}), as well as values of parameters $\eta=1$ and $\theta=\pi$. It is worth noting that the ratio between the initial energy of the gamma quantum and the characteristic energy $\omega_{BW}$ determines the value of parameter $\varepsilon_{2BW}$ (\ref{29}), which can be either greater or less than unity. This significantly affects the number of photons absorbed in the EFBWP. Specifically, if the initial energy of the gamma quantum is less than the characteristic energy, then from equations (\ref{29}) and (\ref{30}) it follows that this process occurs if the number of absorbed wave photons is above a certain minimum $r_{min}$ value, which is greater than unity:
\begin{equation}\label{31}
r\ge r_{min}=\lceil\varepsilon_{2BW}^{-1}\rceil\quad(\omega_2<\omega_{BW}).
\end{equation}
If the initial energy of the gamma quantum is greater than the characteristic energy, then this process takes place already when one photon of the wave is absorbed:
\begin{equation}\label{32}
r\ge1\quad(\omega_2\ge\omega_{BW}).
\end{equation}
Thus, the resonant energy of a positron (for Channel A) or an electron (for Channel B) is determined by two parameters: the corresponding outgoing angle of the positron ($\delta^2_{2+}$) or electron ($\delta^2_{2-}$), and the parameter $\varepsilon_{2BW(r)}$. At the same time, with a fixed parameter $\varepsilon_{2BW(r)}$, for each outgoing angle of the positron or electron, there are two possible energies (see equation (\ref{26})).
\begin{figure}[H]
\begin{minipage}[h]{0.325\linewidth}
\center{\includegraphics[width=0.99\linewidth]{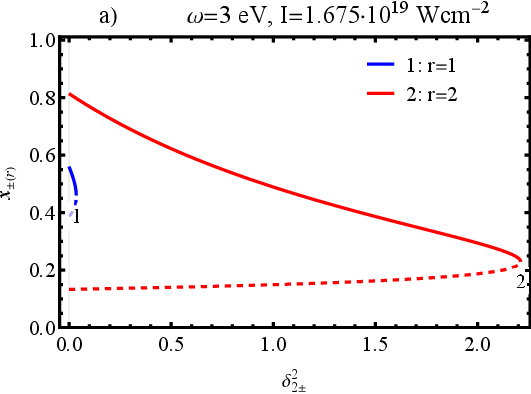}}
\end{minipage}
\hfill
\begin{minipage}[h]{0.325\linewidth}
\center{\includegraphics[width=0.99\linewidth]{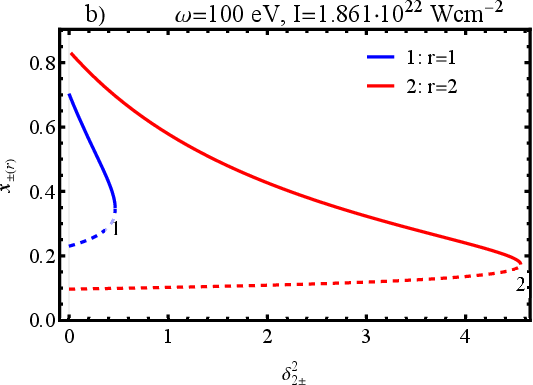}}
\end{minipage}
\hfill
\begin{minipage}[h]{0.325\linewidth}
\center{\includegraphics[width=0.99\linewidth]{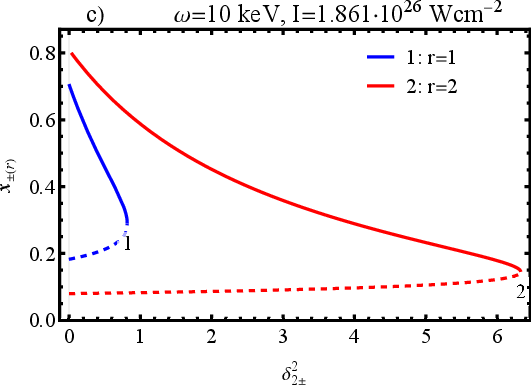}}
\end{minipage}
\caption{The energy of the positron (Channel A) or electron (Channel B) (\ref{26})-(\ref{29}) for the external field-stimulated Breit-Wheeler process with absorption of one and two photons of the wave at different frequencies and intensities of the electromagnetic wave (\ref{30}). Solid lines correspond to the "+"\, and dashed lines correspond to the "--"\, signs before the square root in (\ref{26}). The energies of the initial gamma quanta:  Fig.3a) -- $\omega_1=10\mbox{GeV},\omega_2=180\mbox{GeV}$; Fig.3b) -- $\omega_1=0.5\mbox{GeV},\omega_2=7\mbox{Gev}$; Fig.3c) -- $\omega_1=10\mbox{MeV},\omega_2=80\mbox{MeV}$.}
\label{fig3}
\end{figure}
\par Figure \ref{fig3} shows the dependence of the energy of the positron (for Channel A) or electron (for Channel B) (see equations (\ref{26})-(\ref{29})) for the external field-stimulated Breit-Wheeler process with absorption of one and two photons of the wave at different frequencies, intensities of the electromagnetic wave (equation (\ref{30}), and various initial gamma quanta energies. From this figure, it follows that the interval for the outgoing angle of the positron (electron) significantly depends on the number of absorbed photons of the wave. Additionally, for the same outgoing angle, there are two possible particle energies, except for the maximum outgoing angle.
\par Now let's determine the resonant electron (positron) energy at the first vertex (see Fig. \ref{fig2}). Taking into account equations (\ref{2}), (\ref{3}), (\ref{21}), and (\ref{24}), from the conservation law of the 4-momentum (equation (\ref{23}) of the external field-stimulated Compton effect, we obtain the resonant energies of the electron (for Channel A) or the positron (for Channel B) in terms of the total energy of the initial gamma quanta:
\begin{equation}\label{33}
     x_{\mp(r')}=\frac{\omega_1}{2\omega_i(\varepsilon_{1C(r')}-\delta^2_{1\mp})}\left[\varepsilon_{1C(r')}+\sqrt{\varepsilon_{1C(r')}^2+4(\varepsilon_{1C(r')}-\delta^2_{1\mp})}\right].
\end{equation}
Here is denoted:
\begin{equation}\label{34}
    x_{\mp(r')}=\frac{E_{\mp}(r')}{\omega_i},\quad\delta_{1\mp}=\frac{\omega_1}{m_*}\theta_{1\mp}.
\end{equation}
\begin{equation}\label{35}
\quad\varepsilon_{1C(r')}=r'\varepsilon_{1C},\quad\varepsilon_{1C}=\frac{\omega_1}{\omega_{C}},\quad\omega_{C}=\frac{1}{4}\omega_{BW}.
\end{equation}
Here $\omega_C$ s the characteristic quantum energy of the external field-stimulated Compton effect. This energy is four times less than the characteristic energy for the external field-stimulated Breit-Wheeler process. Additionally, it should be noted that the ultrarelativistic parameter $\delta^2_{1\mp}$, which determines the outgoing angle of the electron or positron, should not take values close to $\varepsilon_{1C(r')}$, in order to satisfy the condition $x_{\mp(r')}<1$ (see equation (\ref{33})). It should also be noted that there are no limitations on the parameter $\varepsilon_{1C(r')}$ for the external field-stimulated Compton effect. Therefore, this process occurs for any number of emitted photons of the wave $r'\ge1$.
\par Furthermore, we will assume that the energies of the initial gamma quanta, within the framework of conditions (\ref{2}), satisfy the additional conditions:
\begin{equation}\label{36}
\omega_2>\omega_{BW},\quad\omega_1\ll\omega_{BW}.
\end{equation}
Conditions (\ref{36}) mean that parameter $\varepsilon_{2BW}>1$, and parameter $\varepsilon_{1BW}\ll1$ (see equations (\ref{29}) and (\ref{30})). Therefore, in Channels A and B, the external field-stimulated Breit-Wheeler process occurs with a number of absorbed photons of the wave $r\ge1$, and for the exchange resonant diagrams A' and B' the number of absorbed photons is $r\ge r_{min}=\lceil\varepsilon_{1BW}^{-1}\rceil\gg1$. Thus, within the framework of conditions (\ref{36}), resonant Channels A' and B' will be suppressed, and we will only consider two resonant Channels A and B (see Fig. \ref{2}). It is also important to consider that for Channel A, the resonant energy of the positron is determined by its outgoing angle relative to the momentum of the second gamma quantum in the EFBWP, while the resonant energy of the electron is determined by its outgoing angle relative to the momentum of the first gamma quantum in the EFSCE. For Channel B, we have the opposite situation, where the energy of the electron is determined by its outgoing angle relative to the momentum of the second gamma quantum, and the energy of the positron is determined by its outgoing angle relative to the momentum of the first gamma quantum (see Fig. \ref{fig2}). Therefore, Channels A and B are distinguishable and do not interfere with each other.
\par
\begin{figure}[H]
\begin{minipage}[h]{0.47\linewidth}
\center{\includegraphics[width=1\linewidth]{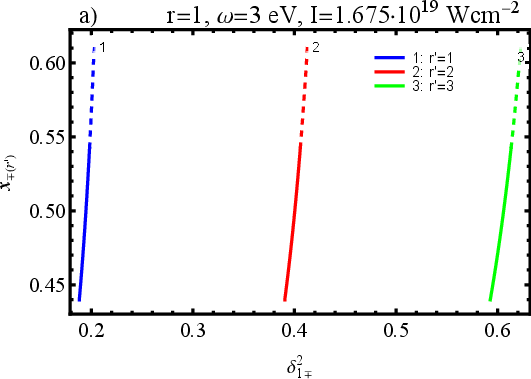}}\\
\end{minipage}
\hfill
\begin{minipage}[h]{0.47\linewidth}
\center{\includegraphics[width=1\linewidth]{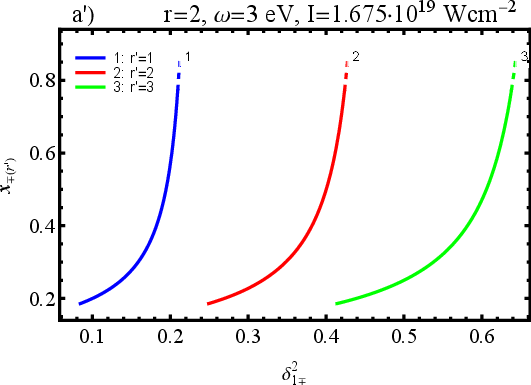}}\\
\end{minipage}
\vfill
\begin{minipage}[h]{0.47\linewidth}
\center{\includegraphics[width=1\linewidth]{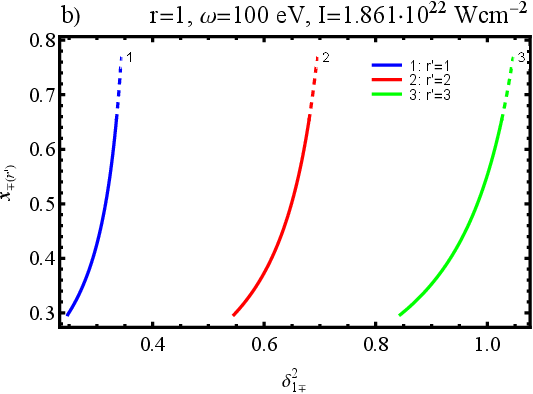}}\\
\end{minipage}
\hfill
\begin{minipage}[h]{0.47\linewidth}
\center{\includegraphics[width=1\linewidth]{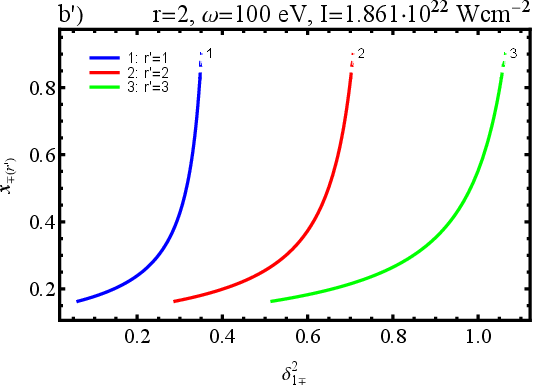}}\\
\end{minipage}
\vfill
\begin{minipage}[h]{0.47\linewidth}
\center{\includegraphics[width=1\linewidth]{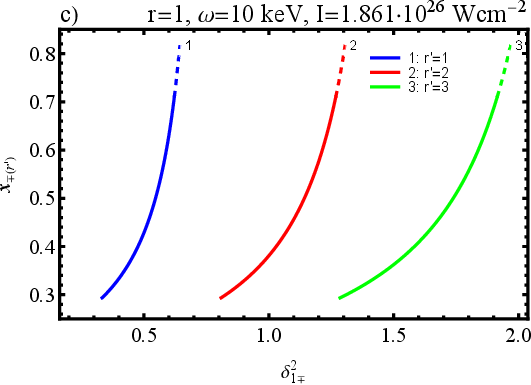}}\\
\end{minipage}
\hfill
\begin{minipage}[h]{0.47\linewidth}
\center{\includegraphics[width=1\linewidth]{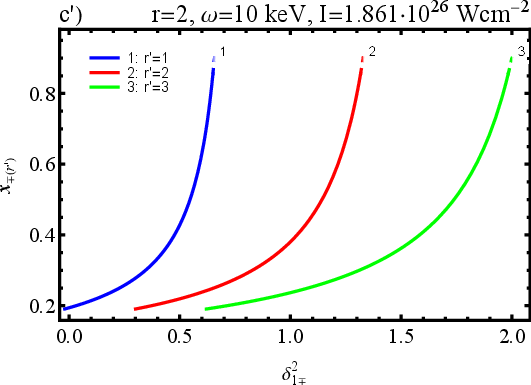}}\\
\end{minipage}
\caption{The dependence of the energy of the electron (Channel A) or positron (Channel B) (\ref{33}), (\ref{38}) for the external field-stimulated Compton effect at different frequencies, intensities of the electromagnetic wave (\ref{30}), and initial gamma quanta energies under the condition of energy conservation in the first and second vertices (\ref{37}). Solid lines correspond to the "+"\, and dashed lines correspond to the "--"\, signs before the square root in expressions (\ref{26}), (\ref{33}), (\ref{38}). The energies of the initial gamma quanta:  Fig.4a), 4a') -- $\omega_1=10\mbox{GeV},\omega_2=180\mbox{GeV}$; Fig.4b), 4b') -- $\omega_1=0.5\mbox{GeV},\omega_2=7\mbox{Gev}$; Fig.4c), 4c') -- $\omega_1=10\mbox{MeV},\omega_2=80\mbox{MeV}$.}
\label{fig4}
\end{figure}

 It is important to note that under resonance conditions (\ref{21}), the resonant energies of the positron and electron for each reaction channel are determined by different physical processes: the external field-stimulated Breit-Wheeler process (\ref{26}) and the Compton external field-stimulated effect (\ref{33}).  At the same time, the energies of the electron-positron pair are related to each other by the general law of conservation of energy
\begin{equation}\label{37}
    x_++x_-\approx1\quad (x_{\pm}=\frac{E_{\pm}}{\omega_i}).
\end{equation}
It should be noted that in equation (\ref{37}) we have neglected a small correction term $|l|\omega/\omega_i\ll1$. Taking into account equations (\ref{26}) and (\ref{33}), as well as the law of conservation of energy (\ref{37}) for Channels A and B, we obtain the following equations relating the outgoing angles of the positron and electron:
\begin{equation}\label{38}
\delta^2_{1\mp}=\varepsilon_{1C(r')}-\frac{(\omega_1/\omega_i)}{(1-x_{\pm(r)})}\left[\varepsilon_{1C(r')}+\frac{(\omega_1/\omega_i)}{(1-x_{\pm(r)})}\right].
\end{equation}
Here the upper (lower) sign corresponds to Channel A (B). In equation (\ref{38}), the left side represents the ultrarelativistic parameter associated with the outgoing angle of the electron (positron) relative to the momentum of the first gamma quantum, and the right side is the function of the ultrarelativistic parameter $\delta_{2\pm}$, associated with the outgoing angle of the positron (electron) relative to the momentum of the second gamma quantum. Under given parameters $\varepsilon_{1C(r')}$ and $\varepsilon_{2BW(r)}$, equation (\ref{38}) uniquely determines the outgoing angles of the electron and positron, and therefore their resonant energies (see Fig. \ref{fig3} and Fig. \ref{fig4}).
\par Figure \ref{fig4} presents the dependence of the energy of the electron (for channel A) or positron (for channel B) (\ref{33}), (\ref{38}) for the external field-stimulated Compton effect at different frequencies, intensities of the electromagnetic wave (\ref{30}), and initial gamma quanta energies under the condition of energy conservation in the first and second vertices (\ref{37}). The graphs are given for different numbers of absorbed ($r$) and emitted ($r'$) photons of the wave.
\par It is also worth noting the important case when the quantum parameter $\varepsilon_{2BW(r)}\gg1$. In this case, from the expression (\ref{26}) with the "+"\, sign before the square root, the energy of the positron (Channel A) or electron (Channel B) approaches the energy of the highly energetic second gamma quantum:
\begin{equation}\label{39}
E_{\pm}\approx\omega_2\left[1-\frac{(1+\delta^2_{2\pm})}{4\varepsilon_{2BW(r)}}\right]\longrightarrow\omega_2\quad(\delta^2_{2\pm}\ll\varepsilon_{2BW(r)}).
\end{equation}
The expression with the "--"\, sign before the square root in equation (\ref{26}) leads to the minimum energy of the positron or electron $E_{\pm}\sim\omega_2/\varepsilon_{2BW(r)}\ll\omega_2$. However, this case is unlikely. Similarly, for the first gamma quantum, when the quantum parameter $\varepsilon_{1C(r')}\gg1$, we obtain the energy of the electron (Channel A) or positron (Channel B) approaching the energy of the first gamma quantum:
\begin{equation}\label{40}
E_{\mp}\approx\omega_1\left[1-\frac{(1+\delta^2_{1\mp})}{\varepsilon_{1C(r')}}\right]\longrightarrow\omega_1\quad(\delta^2_{1\mp}\ll\varepsilon_{1C(r')}).
\end{equation}
Thus, if the quantum parameters $\varepsilon_{1C(r')}$ and $\varepsilon_{2BW(r)}$ take large values, the resonant energies of the positron and electron tend towards the energies of the corresponding initial gamma quanta.

\section{The resonant differential cross section}
\par Previously, it has been shown that in conditions (\ref{2}), (\ref{3}), and (\ref{36}), exchange Channels A' and B' are suppressed. In addition, Channels A and B are distinguishable and therefore do not interfere (see text after equation (\ref{36})). It is also important to note that resonance processes with different numbers of absorbed and emitted wave photons correspond to significantly different probabilities and energies of electron-positron pair. Therefore, they do not interfere either. Due to this, summation over all possible processes with absorption of $r$ wave photons is not necessary in the amplitude (\ref{12}):
\begin{equation}\label{41}
M_{rr'}=\varepsilon_{1\mu}\varepsilon_{2\nu}K^{\mu}_{r'}(\widetilde{p}_-,\widetilde{q}_-)\frac{\hat{q}_-+m}{\widetilde{q}_-^2-m_*^2}K^{\nu}_{-r}(\widetilde{q}_-,-\widetilde{p}_+),\quad r'=l+r.
\end{equation} 
\par The resonant differential cross section for Channels A and B and unpolarized initial gamma quanta and the final electron-positron pair is obtained from the amplitude (\ref{10}), (\ref{11}), (\ref{41}) in a standard way \cite{55}. After simple calculations, we obtain:
\begin{equation}\label{42}
    d\sigma_{rr'}=\frac{2m^6r^2_e}{\widetilde{E}_-\widetilde{E}_+m_*^2\delta^2_{\eta i}}\frac{K_{1\mp(r')}P_{2\pm(r)}}{|\widetilde{q}_{\mp}^2-m_*^2|^2}\delta^{(4)}\left[k_1+k_2-\widetilde{p}_--\widetilde{p}_+-(r'-r)k\right]d^3\widetilde{p}_-d^3\widetilde{p}_+.
\end{equation}
Here, the upper (lower) sign corresponds to Channel A (B), $r_e=e^2/m$ is the classical electron radius. In obtaining the resonant differential cross-section (\ref{42}), the resonant probability was divided by the flux density of the initial gamma quanta \cite{55}:
\begin{equation}\label{43}
    j=\frac{(k_1k_2)}{\omega_1\omega_2}\approx\frac{m^2_*}{2\omega_1\omega_2}\delta^2_{\eta i},\quad\delta^2_{\eta i}\equiv\frac{\omega_1\omega_2}{m^2_*}\theta^2_{i}.
\end{equation}
In expression (\ref{42}), the function $P_{2\pm(r)}$ determines the probability of the external field-stimulated Breit-Wheeler process \cite{1}, and the function $K_{1\mp(r')}$ determines the probability of the external field-stimulated Compton effect \cite{1}:
\begin{equation}\label{45}
    P_{2\pm(r)}=J^2_{r}(\gamma_{2\pm(r)})+\eta^2(2u_{2\pm(r)}-1)\left[\left(\frac{r^2}{\gamma_{2\pm(r)}^2}-1\right)J^2_{r}+J'^2_{r}\right],
\end{equation}
\begin{equation}\label{46}
    K_{1\mp(r')}=-4J^2_{r'}(\gamma_{1\mp(r')})+\eta^2\left[2+\frac{u_{1\mp(r')}^2}{1+u_{1\mp(r')}}\right](J^2_{r'-1}+J^2_{r'+1}-2J^2_{r'}).
\end{equation}
The arguments of the Bessel functions for the external field-stimulated Breit-Wheeler process (\ref{45}) and the external field-stimulated Compton effect (\ref{46}) have the following form:
\begin{equation}\label{47}
    \gamma_{2\pm(r)}=2r\frac{\eta}{\sqrt{1+\eta^2}}\sqrt{\frac{u_{2\pm(r)}}{v_{2\pm(r)}}\left(1-\frac{u_{2\pm(r)}}{v_{2\pm(r)}}\right)},
\end{equation}
\begin{equation}\label{48}
    \gamma_{1\mp(r')}=2r'\frac{\eta}{\sqrt{1+\eta^2}}\sqrt{\frac{u_{1\mp(r')}}{v_{1\mp(r')}}\left(1-\frac{u_{1\mp(r')}}{v_{1\mp(r')}}\right)}.
\end{equation}
Here, the relativistic-invariant parameters are equal to:
\begin{equation}\label{49}
    u_{1\mp(r')}=\frac{(k_1k)}{(p_{\mp}k)}\approx\frac{(\omega_1/\omega_i)}{x_{\mp(r')}},\quad v_{1\mp(r')}=\frac{2r'(q_{\mp}k)}{m^2_*}\approx\varepsilon_{1C(r')}\left(\frac{x_{\mp(r')}}{(\omega_1/\omega_i)}-1\right),
\end{equation}
\begin{equation}\label{50}
    u_{2\pm(r)}=\frac{(k_2k)^2}{4(p_{\pm}k)(q_{\mp}k)}\approx\frac{(\omega_2/\omega_i)}{4x_{\pm(r)}\left(1-\frac{x_{\pm(r)}}{(\omega_2/\omega_i)}\right)},\quad v_{2\pm(r)}=r\frac{(k_2k)}{2m_*^2}\approx\varepsilon_{2BW(r)}.
\end{equation}
\par The elimination of resonant singularity in expression (\ref{42}) is carried out by the Breit-Wigner procedure \cite{49, 54}:
\begin{equation}\label{51}
m_*\longrightarrow\mu_*=m_*-i\Gamma_{\mp(r)},\quad\Gamma_{\mp(r)}=\frac{\widetilde{q}_{\mp}^0}{2m_*}W_1,
\end{equation}
where $W_1$ is the total probability (per unit of time) of the external field-stimulated Compton effect on the intermediate electron (for Channel A) or positron (for Channel B). 
\begin{equation}\label{52}
    W_1=\frac{\alpha m^2}{4\pi\widetilde{q}_{\mp}^0}K(\varepsilon_{1C}),
\end{equation}
\begin{equation}\label{53}
K(\varepsilon_{1C})=\sum_{s=1}^{\infty}\int_0^{s\varepsilon_{1C}}\frac{du}{(1+u)^2}K(u,s\varepsilon_{1C}).
\end{equation}
Here, $\alpha$ is the fine-structure constant, and the function $K(u,s\varepsilon_{1C})$ is determined by the expression:
\begin{equation}\label{54}
    K(u,s\varepsilon_{1C})=-4J^2_{s}(\gamma_{1(s)})+\eta^2\left[2+\frac{u^2}{1+u}\right](J^2_{s-1}+J^2_{s+1}-2J^2_{s})
\end{equation}
\begin{equation}\label{55}
    \gamma_{1(s)}=2s\frac{\eta}{\sqrt{1+\eta^2}}\sqrt{\frac{u}{s\varepsilon_{1C}}\left(1-\frac{u}{s\varepsilon_{1C}}\right)}.
\end{equation}
\par Taking into account the relations (\ref{51})-(\ref{55}), the resonant denominator in the cross-section (\ref{42}) takes the following form:
\begin{equation}\label{56}
|\widetilde{q}_{\mp}^2-m_*^2|^2\longrightarrow m_*^4\frac{x^2_{\mp(r')}}{(\omega_1/\omega_i)^2}\left[\left(\delta^2_{1\mp(0)}-\delta^2_{1\mp}\right)^2+\Upsilon^2_{\mp(r')}\right].
\end{equation}
Here, the ultrarelativistic parameter $\delta^2_{1\mp}$ is related to the resonance energy of the electron (for Channel A) or positron (for Channel B) by the relation (\ref{33}), and the corresponding parameter $\delta^2_{1\mp(0)}$ can take arbitrary values unrelated to the energy of the electron (positron). In this case, the corresponding angular width of the resonance $\Upsilon_{\mp(r')}$ is determined by the expression:
\begin{equation}\label{57}
\Upsilon_{\mp(r')}=\frac{\alpha m^2}{4\pi m_*^2}\frac{\omega_1}{\omega_ix_{\mp(r')}}K(\varepsilon_{1C}).
\end{equation}
Considering relation (\ref{25}), we can set $d^3\widetilde{p}_{\pm}\approx d^3p_{\pm}$ and integrate the three-dimensional momentum of the electron (positron) as well as the energy of the positron (electron) for Channel A (for Channel B) using the delta-function in expression (\ref{42}). After simple calculations, we obtain the following expression for the resonant differential cross-section for Channels A and B:
\begin{equation}\label{58}
    R_{2\pm(rr')}=\frac{d\sigma_{rr'}}{d\delta^2_{2\pm}}=8\pi r^2_e\left(\frac{m}{\delta_{\eta i}\omega_i}\right)^2\frac{x_{\pm(r)}}{x_{\mp(r')}^3}\left(\frac{m}{m_*}\right)^4\left(\frac{\omega_1}{\omega_2}\right)^2\frac{K_{1\mp(r')}P_{2\pm(r)}}{\left[\left(\delta^2_{1\mp(0)}-\delta^2_{1\mp}\right)^2+\Upsilon^2_{\mp(r')}\right]}.
\end{equation}
Here, the upper (lower) sign corresponds to Channel A (B). It should be noted that the differential cross-section (\ref{58}) has a characteristic Breit-Wigner resonance structure \cite{54}. Let's determine the maximum resonant differential cross-section when
\begin{equation}\label{59}
\left(\delta^2_{1\mp(0)}-\delta^2_{1\mp}\right)^2\ll\Upsilon^2_{\mp(r')}.
\end{equation}
Under conditions (\ref{59}), the resonant cross-section (\ref{58}) takes its maximum value, which is equal to:
\begin{equation}\label{60}
R_{2\pm(rr')}^{max}=\frac{d\sigma_{rr'}^{max}}{d\delta^2_{2\pm}}=r^2_ec_{\eta i}\Psi_{\pm(rr')}.
\end{equation}
Here, the function $c_{\eta i}$ is determined by the initial setup parameters
\begin{equation}\label{61}
c_{\eta i}=\frac{2(4\pi)^3}{\alpha^2K^2(\varepsilon_{1C})}\left(\frac{m}{\delta_{\eta i}\omega_2}\right)^2\sim10^8\left(\frac{m}{\delta_{\eta i}\omega_2}\right)^2,
\end{equation}
and the functions $\Psi_{\pm(rr')}$ determine the spectral-angular distribution of the generated electron-positron pair:
\begin{equation}\label{62}
    \Psi_{\pm(rr')}=\frac{x_{\pm(r)}}{1-x_{\pm(r)}}K_{1\mp(r')}P_{2\pm(r)}.
\end{equation}
It is important to emphasize that the magnitude of the maximum resonant differential cross-section significantly depends on the value of the function $c_{\eta i}$ (\ref{61}). Let's require that the function $c_{\eta i}>1$. Then, from relation (\ref{61}), we obtain a condition on the initial ultrarelativistic parameter $\delta^2_{\eta i}$ (\ref{43}):
\begin{equation}\label{63}
\delta^2_{\eta i}<\left(10^4\frac{m}{\omega_2}\right)^2.
\end{equation}
It should be noted that the corresponding Breit-Wheeler differential cross-section without an external field in this kinematics (\ref{24}) has the following order of magnitude \cite{37}:
\begin{equation}\label{64}
\frac{d\sigma_{BW}}{d\delta^2_{2\pm}}\sim r_e^2\left(\frac{m}{\delta_i\omega_i}\right)^2,\quad \delta_i=\frac{\sqrt{\omega_1\omega_2}\theta_i}{m}.
\end{equation}
From relations (\ref{60})-(\ref{62}) and (\ref{64}), it can be seen that the maximum resonant cross-section significantly exceeds the corresponding Breit-Wheeler cross-section without an external field.
\begin{figure}[H]
\begin{minipage}[h]{0.47\linewidth}
\center{\includegraphics[width=1\linewidth]{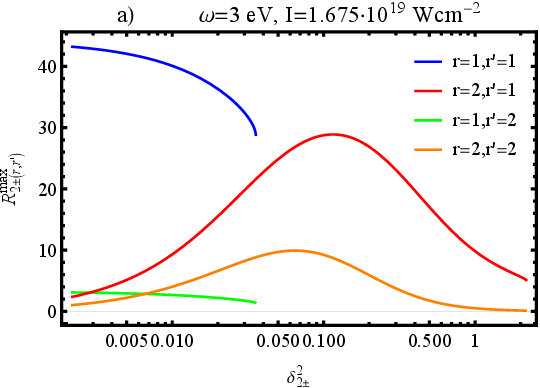}}\\
\end{minipage}
\hfill
\begin{minipage}[h]{0.47\linewidth}
\center{\includegraphics[width=1\linewidth]{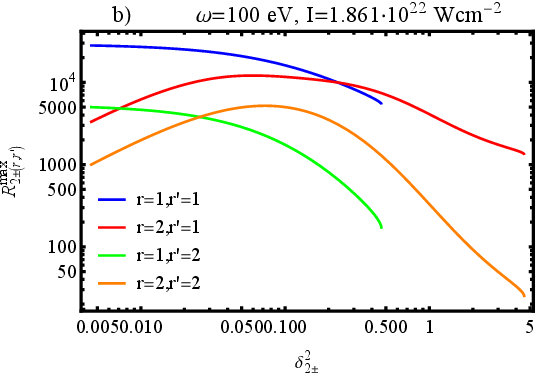}}\\
\end{minipage}
\vfill
\begin{minipage}[h]{0.47\linewidth}
\center{\includegraphics[width=1\linewidth]{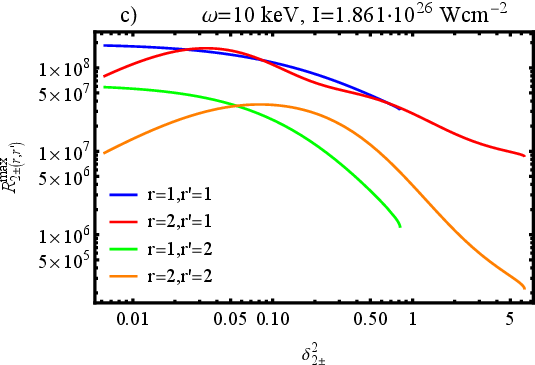}}\\
\end{minipage}
\hfill
\begin{minipage}[h]{0.47\linewidth}
\center{\includegraphics[width=1\linewidth]{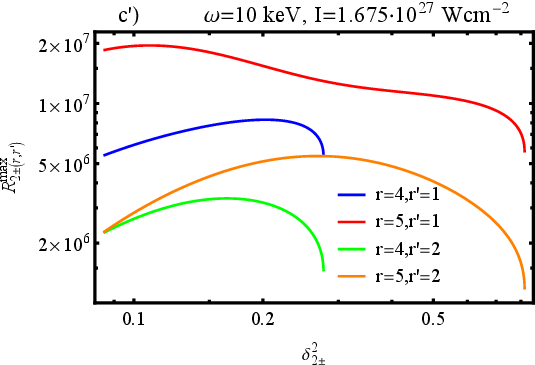}}\\
\end{minipage}
\caption{The dependence of the maximum resonance differential cross-section (\ref{60}) (in units of $r_e^2$) on the positron outgoing angle (Channel A) or electron outgoing angle (Channel B) for various frequencies and intensities, as well as the numbers of absorbed ($r$) and emitted ($r'$) photons. The value of the initial ultrarelativistic parameter $\delta^2_{\eta i}=10^{-4}$. The energies of the initial gamma quanta:  Fig.5a) -- $\omega_1=10\mbox{GeV},\omega_2=180\mbox{GeV}$; Fig.5b) -- $\omega_1=0.5\mbox{GeV},\omega_2=7\mbox{Gev}$; Fig.5c), 5c') -- $\omega_1=10\mbox{MeV},\omega_2=80\mbox{MeV}$.}
\label{fig5}
\end{figure}
\par Figure \ref{fig5} shows the dependencies of the maximum resonance differential cross-section (\ref{60}) on the positron outgoing angle (for Channel A) or electron outgoing angle (for Channel B) for various frequencies and intensities, as well as the numbers of absorbed and emitted photons at the first and second vertices (see Fig. \ref{fig2}). The study focused on the regions of optical and X-ray frequencies of the external strong electromagnetic wave at different sufficiently high energies of initial gamma quanta. It is important to note that the energy of the second high-energy gamma quantum for each frequency and intensity of the wave was chosen according to condition (\ref{32}), in order for the stimulated Breit-Wheeler process to occur with the highest probability, and the energy of the first gamma quantum was chosen to be much lower than the energy of the second gamma quantum (\ref{36}). In this case, with increasing frequency of the external field, the characteristic energy of the Breit-Wheeler process decreased (see relation (\ref{30})). Therefore, energies of initial gamma quanta were chosen to be lower for the X-ray frequency range than for the optical frequency range. As a result, the function (\ref{61}) increased, leading to an increase in the maximum resonance cross-section. This case is shown in Figures \ref{fig5}a) to \ref{fig5}c). However, if the energy of initial gamma quanta remains constant and the intensity of the external field increases, then the maximum resonance cross-section decreases (see Figures \ref{fig5}c) and \ref{fig5}c')). Table 1 displays the values of positron (for Channel A) and electron (for Channel B) energies, as well as the corresponding maximum values of the resonance differential cross-section according to their spectral-angular distribution (see Figures \ref{fig5}a) to \ref{fig5}c')) for different frequencies and intensities of the wave, as well as different energies of initial the gamma quanta.
\begin{table}[H]
    \centering
    \begin{tabular}{|c|c|c|c|c|}
         \hline
                  &($r,r'$)&$\delta^2_{2\pm max}$&$x_{\pm(r)}$&$R^{max}_{2\pm(rr')}$\\ \hline
\multirow{4}{*}{\begin{tabular}{c}
$I=1.675\cdot10^{19}\mbox{Wcm}^{-2}$,\\
$\omega=3\mbox{eV}$,\\
$\omega_1=10\mbox{GeV}$,\\
$\omega_2=180\mbox{GeV}$\end{tabular}} & (1,1)       &  0 & \multirow{2}{*}{0.56} & 44  \\ \cline{2-3} \cline{5-5} 
                  &    (1,2)    &  0 &  & 3  \\ \cline{2-5} 
                  &   (2,1)     & 0.12  &  0.76                 & 29  \\ \cline{2-5} 
                  &  (2,2)      & 0.07  &  0.78                 &10   \\ \hline
\multirow{4}{*}{\begin{tabular}{c}
$I=1.861\cdot10^{22}\mbox{Wcm}^{-2}$,\\
$\omega=100\mbox{eV}$,\\
$\omega_1=0.5\mbox{GeV}$,\\
$\omega_2=7\mbox{GeV}$\end{tabular}} &  (1,1)      & 0  & \multirow{2}{*}{0.7} & $2.9\cdot10^4$  \\ \cline{2-3} \cline{5-5} 
                  &  (1,2)      & 0  &                   &  $5.3\cdot10^3$ \\ \cline{2-5} 
                  &   (2,1)     & 0.06  &  0.82                 &  $1.2\cdot10^4$ \\ \cline{2-5} 
                  &   (2,2)     & 0.07  &    0.81               & $5.2\cdot10^3$  \\ \hline
\multirow{4}{*}{\begin{tabular}{c}
$I=1.861\cdot10^{26}\mbox{Wcm}^{-2}$,\\
$\omega=10\mbox{keV}$,\\
$\omega_1=10\mbox{MeV}$,\\
$\omega_2=80\mbox{MeV}$\end{tabular}} &  (1,1)      &  0 & \multirow{2}{*}{0.71} &$1.9\cdot10^8$   \\ \cline{2-3} \cline{5-5} 
                  &   (1,2)     & 0  &                   & $6.3\cdot10^7$  \\ \cline{2-5} 
                  &   (2,1)     & 0.03  &    0.8               &  $1.7\cdot10^8$ \\ \cline{2-5} 
                  &  (2,2)      &  0.08 &  0.79                 &  $3.6\cdot10^7$ \\ \hline
\multirow{4}{*}{\begin{tabular}{c}
$I=1.675\cdot10^{27}\mbox{Wcm}^{-2}$,\\
$\omega=10\mbox{keV}$,\\
$\omega_1=10\mbox{MeV}$,\\
$\omega_2=80\mbox{MeV}$\end{tabular}} &  (4,1)      & 0.2  &         0.47          &  $8.3\cdot10^6$ \\ \cline{2-5} 
                  &  (4,2)     & 0.17 &    0.5                &  $3.4\cdot10^6$ \\ \cline{2-5} 
                  &    (5,1)    &0.11   &   0.64                &  $2\cdot10^7$ \\ \cline{2-5} 
                  & (5,2)       &  0.27 &   0.56                & $5.5\cdot10^6$  \\ \hline
    \end{tabular}
    \caption{The maximum
values of the resonance differential cross-section.}
    \label{tab1}
\end{table}
\par From Table 1, it can be observed that if the energy of one of the initial gamma quanta slightly exceeds the characteristic Breit-Wheeler energy, the production of electron-positron pairs occurs with a very large cross-section. For the optical frequency range, the resonance differential cross-section can exceed the value in magnitude by a factor of 44, while for the X-ray frequency range, it can exceed the value by eight orders of magnitude. In this case, the positrons (electrons) are emitted in a narrow cone and with very high energy.
\section*{Conclusion}
\par We considered the resonant Breit-Wheeler process modified by an external strong electromagnetic field for high-energy initial gamma quanta when the energy of one of them significantly exceeded the energy of the other. The following results were obtained:
\begin{enumerate}
    \item The resonant kinematics of the process has been studied in detail. It was demonstrated that the problem involves two characteristic energies: the Breit-Wheeler energy $\omega_{BW}$ (\ref{30}) and the Compton effect energy $\omega_C$ (\ref{35}). These energies differ from each other by a factor of four. The ratios of the initial gamma quanta energies to these characteristic energies significantly affect the number of absorbed or emitted wave photons and, ultimately, the probability of the process.
    \item The resonant energies of the positron and electron strongly depend on their outgoing angles, as well as the characteristic quantum parameters $\varepsilon_{2BW(r)}$ (\ref{29}) and $\varepsilon_{1C(r')}$ (\ref{35}). Furthermore, the outgoing angles of the electron and positron are interdependent (\ref{38}).
    \item The maximum resonant differential cross-section is achieved when the energy of one of the initial gamma quantum slightly exceeds the characteristic Breit-Wheeler energy. In this case, for the optical frequency range and $\omega_2=180\mbox{GeV}$, the maximum resonant cross-section is $R_{2\pm(rr')}^{max}=44r^2_e$, whereas for the X-ray frequency range, it is $R_{2\pm(rr')}^{max}\sim(10^6\div10^8)r^2_e$.
\end{enumerate}
\par The obtained results can be utilized to achieve ultrarelativistic positron (electron) beams with a very high probability in the external field-modified Breit-Wheeler process. Additionally, these results can be employed to explain the fluxes of ultrarelativistic positrons (electrons) near neutron stars and magnetars, as well as in the modeling of physical processes in laser-induced thermonuclear fusion.
\par The research was funded by the Ministry of Science and Higher Education of the Russian Federation under the strategic academic leadership program “Priority 2030” (Agreement 075-15-2023-380 dated 20.02.2023).

\end{document}